\documentclass[pre,twocolumn,letterpaper,floatfix,showpacs]{revtex4}
\usepackage{color,graphicx,amsmath,amssymb,amsfonts}
\DeclareGraphicsExtensions{.ps,.eps}

\begin{document}
\title{ Generating Function for Particle-Number Probability
       Distribution in Directed Percolation} \author{Lucian Anton}
       \email{lucian.anton@manchester.ac.uk} \affiliation{Department
       of Physics and Astronomy, University of Manchester, M13 9PL,
       U.K.}  \affiliation{Institute of Atomic Physics, INFLPR, Lab
       22, PO Box MG-36 R76900, Bucharest, Romania} \author{Hyunggyu
       Park} \email{hgpark@kias.re.kr} \affiliation{School of Physics,
       Korea Institute for Advanced Study, Seoul 130-722, Korea}
       \author{Su-Chan Park} \email{psc@kias.re.kr}
       \affiliation{School of Physics, Korea Institute for Advanced
       Study, Seoul 130-722, Korea}

\date{\today} 

\begin{abstract}
We derive a generic expression for the generating function (GF) of the
particle-number probability distribution (PNPD) for a simple reaction
diffusion model that belongs to the directed percolation universality
class. Starting with a single particle on a lattice, we show that the
GF of the PNPD can be written as an infinite series of cumulants taken
at zero momentum. This series can be summed up into a complete form at
the level of a mean-field approximation.  Using the renormalization
group techniques, we determine logarithmic corrections for the GF at
the upper critical dimension. We also find the critical scaling form
for the PNPD and check its universality numerically in one dimension.
The critical scaling function is found to be universal up to two
non-universal metric factors.
\end{abstract}

\maketitle

\section{Introduction} 

In the last decades, the statistical mechanics of non-equilibrium models
characterized by large scale space-time fluctuations and absorbing
phase transition has been  studied extensively with numerical and
analytical tools \cite{Hinrichsen00}.
The directed percolation (DP) universality class covers a large
variety of models of this type, which have short range interactions, a
single absorbing state, and a scalar order parameter \cite{J81,G82}.
The DP conjecture has been extended to models with multiple components
and even multiple absorbing states, unless there is an additional
symmetry and a conservation law. Nevertheless we should mention that
there is at least one exception in the case of more elaborated
interactions \cite{HH,Janssen04a,Park04}.

Besides various numerical techniques, the response functional
formalism combined with the renormalization group (RG) techniques is
one of the powerful analytical tool to explore the asymptotic
properties near criticality for these models
\cite{Cardy98,Cardy99,Tauber05,Janssen00,Janssen05}.  The DP
universality is described by the Reggeon field theory which has been
investigated in details by many researchers \cite{Hinrichsen00}.  Its
upper critical dimension is known to be 4 and the values of the
critical exponents were obtained up to ${\cal O} ({\epsilon}^2)$ with
space dimension $d=4-\epsilon$ \cite{J81,JKO99}. The leading logarithmic
corrections at the upper critical dimension are also available in
the literature \cite{Wil98,Janssen04} and tested numerically
\cite{Lubeck04}.  Recently the same field-theoretical formalism was
successfully applied to identify and calculate the survival
probability and the mean number of particles, which are the central
quantities of interest in the damage-spreading (DS) type dynamics
\cite{Janssen03,Janssen04}.

The aim of this paper is to explore in the same framework a more
general and useful quantity in the DS  type dynamics; the generating
function (GF) of the 
particle-number probability distribution (PNPD).  In principle, all
moments of the number of particles including the survival probability
can be derived from the GF.  We start from a lattice description of
a reaction diffusion model that belongs to the DP universality
class. Then, we use the bosonic operator formalism \cite{Cardy99}
to represent the master equation and the associated field theory to
define the generating function (GF) of the PNPD. With a single
particle initially, we derive the GF of the PNPD as an infinite series
of multi-point Green functions. This series can be summed up
explicitly under the tree (mean-field) approximations into a complete
form.

Assuming the generic critical scaling property of the Green functions,
we obtain the critical scaling form of the PNPD. Using the RG
techniques, we calculate the leading logarithmic corrections for the
critical GF at the upper critical dimension. Our results are
consistent with recent reports for the survival probability and the
average number of particles by Janssen and Stenull
\cite{Janssen04,Janssen03}.

The conventional RG analysis suggests that the critical scaling
function for the PNPD should be universal up to two non-universal
metric factors which  microscopic details of the models are integrated
into. We check this universality numerically in one dimension for two
variants of bosonic DP models and also for the hard core DP
models where the site occupancy of particles is restricted to
one. Numerical simulations show that all the  models form a
universality very nicely at criticality with only two adjusting parameters.

The paper is organized as follows: Section ~\ref{sec:genfor}
introduces the model and defines the GF for the PNPD. 
Secion ~\ref{sec:gf} is devoted to the derivation of 
the series expansion of the GF in terms
of connected Green functions. In Sec.~\ref{sec:mf}, we derive the mean
field solution and present the logarithmic corrections to the scaling
in the critical state at the upper critical dimension for the first
three moments of the PNPD. In Sec.~\ref{sec:scaling}, we analyze the
scaling form of the critical PNPD, using the RG equations and test it
numerically in one dimension for both bosonic models and hard core
models.  We present the conclusion in Sec. \ref{sec:concl} and the
technical details for the calculation of the logarithmic corrections
in the Appendix.

\section{\label{sec:genfor}General formalism}

We consider a reaction diffusion model (the Gribov process)
\cite{J81,G82} that belongs to the DP universality. It is defined on a
$d$-dimensional lattice with $N^d$ sites which may be either vacant or
occupied by  particles.  The evolution rule is
summarized in the stoichiometric notation as

\begin{align}
  A \varnothing &\stackrel{D_l}{\longleftrightarrow} \varnothing A \ ,\notag\\
  A&\stackrel{\lambda_a}{\longrightarrow} \varnothing\ , \notag\\
  2A&\stackrel{\lambda_c}{\longrightarrow} A\ ,\label{eq:tran}\\
  A&\stackrel{\lambda_b}{\longrightarrow} 2A\ , \notag
\end{align}
where $D_l$ is the nearest neighbor hopping (diffusion) rate,
$\lambda_a$ is the annihilation rate for a particle, $\lambda_c$ is
the coagulation rate for two particles at a site,
and $\lambda_b$ is the branching rate.

A microscopic state of the system is defined by the set of particle numbers
at each lattice site: ${\bf n}=\{n_i\}$. The time evolution of 
the probability distribution $P({\bf n},t)$ can be
described by the master equation for the state vector $|\Psi(t)\rangle
=\sum_{\bf n} P({\bf n},t) |{\bf n}\rangle$ at time $t$:
\begin{equation}
  \frac{d|\Psi\rangle}{dt}=-H|\Psi\rangle\ .
\end{equation}
The evolution operator $H$ can be written in terms of 
bosonic operators as
\begin{eqnarray}
    H&=&D_l\sum_{\langle
    ij \rangle}(a^{\dagger}_{i} -a^{\dagger}_{j}) (a_{i}
    -a_{j})-\lambda_a\sum_{i}(a_i-a^{\dagger}_{i}a_{i}) \nonumber \\
    &&-\lambda_c\sum_{i}(a^{\dagger}_{i}a^{2}_{i} -
    {a^{\dagger}}^{2}_{i}a^{2}_{i}) -\lambda_b
    \sum_{i}({a^{\dagger}}^2_{i}a_{i}-{a^{\dagger}}_{i}a_{i}) \
    ,\label{eq:initham}
\end{eqnarray}
where $a_i (a^{\dagger}_i)$ is the annihilation (creation) operator of
bosonic particles, satisfying the commutation relation
$[a_i,a^\dagger_j]=\delta_{ij}$.

The average of a quantity $O$ at time $t$  is computed by the formula
\begin{equation}\label{eq:genav}
  \langle O
  \rangle_t=\langle0|e^{\sum_{i}a_{i}} {\hat O} e^{-tH}   |\Psi(0) \rangle
\end{equation}
for a given initial state $|\Psi(0) \rangle$ \cite{Cardy98,Cardy99,Tauber05},
where $|0\rangle$ represents the vacuum state and ${\hat O}$ is the
operator representation of the quantity $O$. Similarly, one can easily
derive the probability that the system has $n$ particles on the
lattice at time $t$ \cite{Howard98} as
\begin{equation}\label{eq:pntdef}
p(n,t)=\frac{1}{n!} \langle0|\Bigl(\sum_{i}a_{i}\Bigr)^ne^{-tH}
  |\Psi(0) \rangle\ ,  
\end{equation}
and its generating function (GF) is defined as 
\begin{equation}
  F(s,t)=\sum_{n=0} s^np(n,t)=\langle0|e^{\sum_{i}sa_{i}}e^{-tH}
  |\Psi(0) \rangle\ .\label{eq:genfunction1}
\end{equation}

\section{\label{sec:gf} The expansion of the generating function}

It is well known that the calculation of the average 
becomes simpler by moving the factor
$e^{\sum_{i}a_{i}}$ in Eq.~\eqref{eq:genav}  to the rightmost position.  Since
$e^{a}a^\dag=(a^\dag+1)e^a$, this move results in shifting
$a^\dagger\rightarrow a^\dagger +1$ in the operators ${\hat O}$ and
$H$.  We use the same trick to calculate the GF in
Eq.~\eqref{eq:genfunction1}, and then
\begin{equation}
  F(s,t)=\langle0|e^{\sum_{i}(s-1)a_{i}}e^{-tH_s}e^{\sum_{i}a_{i}}
  |\Psi(0) \rangle\ ,\label{eq:genfunction1-1}
\end{equation}
where the shifted Hamiltonian $H_s(a^\dagger, a)=H(a^\dag+1, a)$.

Through the standard coherent-state path integral
formulation~\cite{Cardy99,Tauber05}, one can find the shifted action
$S(\bar\phi, \phi)$ such that the path integral is weighed by $e^{-S}$
as
\begin{align}
& S(\bar\phi,\phi) =\int dt \left[\sum_i \bar\phi_i\partial_t\phi_i
  +H_s(\bar\phi,\phi)\right] \nonumber\\ &= \int dt \left[\sum_i
  \bar\phi_i\partial_t\phi_i + D_l\sum_{\langle
  ij\rangle}(\bar\phi_i-\bar\phi_j) (\phi_i-\phi_j) \right
  .\nonumber\\ & \left .+\!\sum_i
  \left\{(\lambda_a-\lambda_b)\bar\phi_i\phi_i\! +\!
  \lambda_c\bar\phi_i\phi_{i}^2\!-\! \lambda_b{\bar\phi_{i}}^2\phi_{i}\!
  +\!  \lambda_c{\bar\phi_{i}}^2\phi_{i}^2 \right\} \right].
\label{eq:action}
\end{align}
For the calculation of the asymptotic properties, the action is taken
formally in the continuum limit with the transformations
\begin{align}
 & \sum_i\rightarrow b^{-d}\!\int dx^d, \ \
 f_{x_i+be_j}-f_{x_i}\rightarrow b\nabla_{x_j} f(x),\nonumber\\ &
 \phi_i(t)\rightarrow b^{d} \phi(x,t), \quad \bar\phi_i(t) \rightarrow
 \bar\phi(x,t)\ , \\ & D=b^{2}D_l,\ \ \mu=\lambda_a-\lambda_b, \ \
 g'=b^{d} \lambda_c, \ \ g''=\lambda_b, \nonumber
\end{align}
where $b$ is the lattice constant and $e_j$ is the unit vector along the
 $j$ axis.  In the continuum limit, the action reads
\begin{align}\label{eq:continuumaction}
  S=&\int dx dt \left[ \bar\phi(x,t) \partial_t\phi(x,t)
+D(\nabla\bar\phi(x,t)) (\nabla\phi(x,t))\right. \notag\\ &\left.+
\mu\bar\phi(x,t)\phi(x,t)  +\frac{1}{2}
g'\bar\phi(x,t)\phi^2(x,t) \right.\notag\\ &\left.  - \frac{1}{2}
g''{\bar\phi}^2(x,t)\phi(x,t)\right] ,\
\end{align}
where  the term $g'{\bar\phi}^2\phi^2$ is dropped as being irrelevant.
The coupling constants $g'$ and $g''$ can be reduced
to one coupling constant $g=\sqrt{g'g''}$ by rescaling the fields as
follows: $\phi \to\alpha \phi$, $\bar\phi \to \alpha^{-1} \bar\phi$,
with $\alpha=\sqrt{g''/g'}$.  The rescaled ``symmetric'' action is then 
invariant under the time reversal transformation: $t\to -t$,
$\phi\to-\bar\phi$, $\bar\phi\to -\phi$.

Consider the GF with one particle initially: $|\Psi(0)
\rangle=a_0^\dag |0\rangle$.  Then, we have
\begin{align}
  &F(s,t)=\langle0|e^{\sum_i (s-1) a_i}e^{-tH_s}(a^{\dag}_{0}+1)|0
  \rangle \notag\\ &=1+\langle0|e^{\sum_i (s-1) a_i} e^{-tH_s}
  a^{\dag}_{0}|0\rangle
  \label{eq:gf_operator}\nonumber\\ 
&=1+\sum_{p=1}\frac{(s-1)^p}{p!}
\sum_{i_1,\dots,i_p} \langle 0| a_{i_1}(t) \dots a_{i_p}(t) e^{-tH_s}
a^\dag_{0}(0) |0 \rangle\ ,
\end{align}
where we used the relations $H_s|0\rangle=0$ and $\langle 0| H_s$=0
because each term in $H_s$ has $a^\dag$ to the left and $a$ to the right.
In the path-integral formulation, the above correlation functions are simply
connected Green functions (cumulants) with the shifted action in 
Eq.~(\ref{eq:action}). In the continuum limit with the symmetric action, the GF
is written as
\begin{align}
  &F(s,t)=1+\sum_{p=1}\frac{(s-1)^p}{p!} \alpha^{p-1} \notag\\ &
  \times \int dx_1,\dots dx_p G^{(p,1)}(x_1,t,\dots x_p,t;0,0)\ ,
  \label{eq:cumexp-cl}
\end{align}
where 
\begin{align}
G^{(p,1)}(x_1,t,\dots x_p,t;0,0)=\langle \phi(x_1,t)\dots
\phi(x_p,t) \bar\phi(0,0) \rangle
\end{align} are the connected Green functions for the
symmetric action and the factor $\alpha=\sqrt{g''/g'}$ stems from the
rescaling of the fields.

Close to the critical point, we can use the renormalized
Green functions of which the scaling behavior is known. Then, we have
\begin{align}
  & G^{(p,1)}(x_1,t,\dots x_p,t;0,0) \notag\\ &=
   Z^{p/2}_{\phi}Z^{1/2}_{\bar\phi} 
   G^{(p,1)}_R(x_1,t,\dots x_p,t;0,0) \notag\\ &=
   Z^{p/2}_{\phi}Z^{1/2}_{\bar\phi}
   t^{-p\frac{\beta}{\nu_\parallel}-\frac{\beta'}{\nu_\parallel}}
   \mathcal{G}^{(p,1)}(x_1/t^{1/z},\dots x_p/t^{1/z}) \ ,
\end{align}
where $Z_\phi$, $Z_{\bar\phi}$ are the fields renormalization factors
that are equal for the time reversal symmetric theory
and we have used the scaling property of the renormalized Green
functions close to the critical point as
\begin{align}
  & G^{(p,1)}_R(x_1,t_1,\dots x_p,t_p;x_0,t_0)=|\mu|^{p\beta+\beta'}
  \notag\\ & \times H(|\mu|^{\nu_\perp}
  x_1,|\mu|^{\nu_\parallel}t_1,\dots)\ ,
\end{align}
where $\mu$ is the deviation from the critical point, and $\beta$,
$\beta'$, $\nu_\parallel$, $\nu_\perp$ and $z$ are the standard
critical exponents defined in Ref. \cite{Hinrichsen00}.

The generating function can be written now as
\begin{align}
   &\bar{F}(s,t)\equiv F(s,t)-1 \notag\\ &=\sum_{p=1}\frac{(s-1)^p}{p!}
   \alpha^{p-1} Z^{p/2}_{\phi}Z^{1/2}_{\bar\phi}
   t^{-p(\frac{\beta}{\nu_\parallel}
   -\frac{d}{z})-\frac{\beta'}{\nu_\parallel}} A^{(p,1)} \notag\\ &=
   t^{-\frac{\beta'}{\nu_\parallel}} Z^{1/2}_{\phi}Z^{1/2}_{\bar\phi}
   \left[\sum_{p=1}\frac{(s-1)^p}{p!}(\alpha Z_{\phi}^{1/2})^{p-1}
   t^{p(\frac{d}{z}-\frac{\beta}{\nu_\parallel})} A^{(p,1)}\right]\ ,
  \label{eq:cumexp-aplitudes}
\end{align}
where 
\begin{equation}
 A^{(p,1)}=\int dx_1\dots, dx_p \mathcal{G}^{(p,1)}(x_1,\dots,x_p)\ .
\end{equation}
This expression shows that $\bar{F}(s,t)$ can be renormalized by
multiplication with $Z^{-1/2}_{\phi}Z^{-1/2}_{\bar\phi}$ after the
parameter $\alpha$ is renormalized as $\alpha_R =
Z^{1/2}_{\phi}\alpha$ .

\section{\label{sec:mf} Mean field approximation and logarithmic corrections}

\subsection{Mean-field solution}

In Eq.~\eqref{eq:cumexp-aplitudes}, the GF  is
expressed as a formal series whose convergence property depends on the
coefficients $A^{(p,1)}$. As a first step to study this series, we
calculate the mean field (MF) solution for the GF.  In this
approximation, we consider only the tree diagrams for the multi-point
correlation function $G^{(p,1)}$ in Eq.~(\ref{eq:cumexp-cl}).

\begin{figure}
\includegraphics{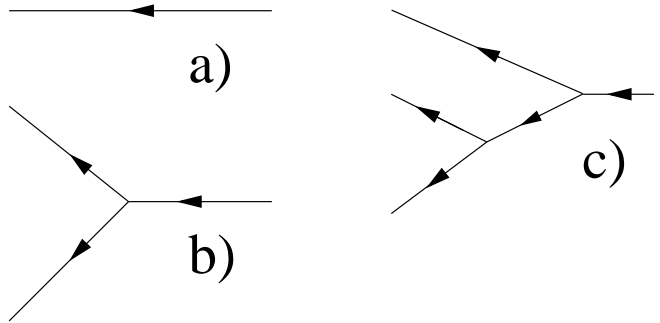}
\caption{\label{fig:treediagrams} The first three tree diagrams used
in the mean-field approximation.}
\end{figure}

The lowest order approximation of the $(p,1)$ Green function can be
obtained by connecting $p-1$ vertices of type $\bar{\phi}^2\phi$ with
free two point Green functions, as shown in
Fig. \ref{fig:treediagrams} where each line is associated with the
free two-point $(1,1)$ Green function in the momentum space as
\begin{equation}
  G(k,t)=\theta(t)e^{-t(Dk^2+\mu)}.
\end{equation}
For the sum of tree diagrams we can easily obtain an integral recurrence
equation for the GF by resuming the diagrams after the first vertex
from the right in the tree, which is
\begin{equation}\label{eq:mf_rec}
  \bar{F}(s,t)=\bar{F}(s,0) e^{-\mu t}+
  \frac{1}{2}g''\int_{0}^tdt'{\bar{F}(s,t-t')}^2G(k=0,t'),
\end{equation}
that is equivalent with the differential  equation
\begin{equation}
  \frac{d}{dt} \bar{F}(s,t)=-\mu \bar{F}(s,t) +
  \frac{1}{2}g''{\bar{F}(s,t)}^2 .
\end{equation}

With the single particle initial condition $F(s,0)=s$ $(\bar{F}(s,0)=s-1)$,
one can  find the exact solution for the GF as
\begin{equation}\label{eq:psmf}
  F(s,t)=1+\frac{(s-1)e^{-\mu t}}{1-(s-1)\frac{g''}{2\mu}(1-e^{-\mu
     t})}\ . 
\end{equation}

By inverting the GF in Eq.~(\ref{eq:genfunction1}), we find the PNPD 
 as
\begin{align}
  p(n,t)=\frac{e^{-\mu t}}{(1+\frac{g''}{2\mu}(1-e^{-\mu t}))^2}
  \left( \frac{\frac{g''}{2\mu}(1-e^{-\mu
  t})}{1+\frac{g''}{2\mu}(1-e^{-\mu t})}\right)^{n-1}, 
\end{align}
for $n\ge 1$. At criticality ($\mu=0$), the PNPD behaves in the
asymptotic limit (large $t$) as
\begin{align}\label{eq:pntmf}
p(n,t)\approx \left(\frac{2}{g''t}\right)^2 e^{-\frac{2(n-1)}{g''t}}.  
\end{align}

The survival probability $p_s(t)$ that the system is active at time
$t$ can be derived  from the  expression
$p_s(t)=1-p(0,t)$$=1-F(0,t)$.  From Eq. \eqref{eq:psmf}, we find
\begin{align}\label{eq:spmf}
  p_s(t)=\frac{e^{-\mu t}}{1+\frac{g''}{2\mu}(1-e^{-\mu
     t})} \ .
\end{align}
We also find the mean number of particles 
$N(t)=\langle n \rangle=\sum_n n p(n,t)=\partial F/\partial s |_{s=1}$ as
\begin{equation}
N(t)=e^{-\mu t}.
\end{equation}
Similarly, all higher moments $\langle n^k \rangle$ can be obtained by
differentiating the GF with respect to $s$ and taking the $s\to 1$
limit.

At criticality, the survival probability decays algebraically as
\begin{equation}\label{eq:spmfc}
   p_s(t)=\frac{1}{1+g''t/2}\approx \frac{2}{g''} t^{-1}\ ,
\end{equation}
while $N(t)$ remains a constant of unity. This is consistent with other
MF  predictions for the dynamic scaling exponents: 
$\delta_{MF}=1$ and $\eta_{MF}=0$ \cite{Hinrichsen00}. 

Near criticality, we find, for the large $t$ limit, that
$p_s (t)\approx (2\mu/g'') e^{-\mu t}$ in the absorbing side $(\mu\gtrsim0)$
and $p_s (t)\approx (2|\mu|/g'')(1+e^{-|\mu|t})$ in the active side 
$(\mu\lesssim 0)$. These results are fully consistent with the recent results
by Janssen \cite{Janssen03}.

\subsection{Logarithmic corrections to critical scaling}

At the critical dimension, where the coupling constant is
 dimensionless, the MF solution \eqref{eq:psmf} can be improved
 systematically at large time with the help of the renormalization
 group equations by adding the one-loop correction.  A
 detailed account of this procedure has been presented recently in
 reference \cite{Janssen04}. In our case we use one-loop approximation
 for the GF, combined with the two-loop Wilson functions taken from
 literature. The details of the calculation for one-loop GF and its
 asymptotic scaling derived with the help of RG equations are
 presented in the Appendix. In this section we present, for the
 critical state, in a parametric form, the leading order expressions
 of the survival probability, $p_s(t)$, and of the first two moments
 of the particle number distribution, $N(t)$, $N_2(t)$, which might be
 of interest for numerical experiments.

From Eq. \eqref{eq:gflog} we have at large $t$
\begin{align}\label{eq:ps_log}
  p_s(t)&=-\bar F(0,t) \notag \\ &= \frac{4N'}{\bar D t}w^{-1}
   e^{c_0w^2+O(w^4)} \left( 1+( 3\ln X_0-9/2) w^2 \right)\notag\\
   &+O(w^{-5/6}t^{-2})\ , \\ N(t)&=\frac{\partial \bar
   F(s,t)}{\partial s}\Big\vert_{s=1}\notag \\ &= N' w^{-1/3} e^{c_1
   w^2+O(w^4)} ( 1 + (\ln X_0 + 1) w^2 ) \ ,\label{eq:nlog}\\ N_2(t)
   &= \frac{\partial^2 \bar F(s,t)}{\partial
   s^2}\Big\vert_{s=1}+N(t)\notag \\ &=\frac{1}{2} N'\bar{D} t w^{1/3}
   e^{c_2 w^2+O(w^4)} \notag\\ & \times ( 1 - (\ln X_0 -4) w^2
   )+O(w^{-1/3}) \ ,
\end{align}
where $w$ depends on time $t$ through Eqs. (\ref{eq:woftau},
\ref{eq:tauoft}).
In the above expressions, the universal constants $c_0$, $c_1$, $c_2$
are defined in terms of the coefficients of the Wilson functions, Eqs
(\ref{eq:beta}- \ref{eq:will2}), and have the values
\begin{align}
  c_0&= \frac{1}{48}\left(7-34\ln\frac{4}{3}\right)\\
  c_1&= -\frac{1}{144}\left(25+322\ln\frac{4}{3}\right)\\
  c_2&= -\frac{1}{144}\left(71+542\ln\frac{4}{3}\right) \ .
\end{align}
  The remaining $N'$, $\bar D$, $X_0$ and $t_0$ are non-universal
constants whose detailed origins can be explored in the Appendix.

We make the observation that Eqs. (\ref{eq:ps_log},\ref{eq:nlog}) are
equivalent with the expressions of Ref. \cite{Janssen04} after the
change of variable $ w\to (w/8)^{1/2}$.

\section{\label{sec:scaling} Critical scaling functions for the PNPD}

Since the GF of the PNPD is renormalizable, see
Eq.~\eqref{eq:cumexp-aplitudes}, we can use the RG equation to find
its scaling properties in the critical state. At large $t$, we can
extract the scaling property of the PNPD through the GF scaling
behavior.

A dimensional analysis with Eq.~\eqref{eq:cumexp-aplitudes} suggests the
following form for the GF
\begin{equation}\label{eq:dim-ps}
  \bar{F}(s,t)=yf(D\kappa^2t, y\alpha \kappa^{-d/2}, u)\ ,
\end{equation}
where $\kappa^{-1}$ is an arbitrary length,
$u$ is the dimensionless coupling constant (see Appendix),
and $y=s-1$. 

The independence of the bare $\bar{F}$ on $\kappa$ yields the
renormalization group equation
\begin{equation}
  \kappa\frac{\partial}{\partial \kappa} \left( yZ_{\phi}f(D\kappa^2t,
  y \alpha \kappa^{-d/2},u) \right)=0
\end{equation}
or, in more  details, we have
\begin{align}\label{eq:callan}
 &\left(\gamma_{\phi}+ \kappa\frac{\partial}{\partial \kappa} +\gamma_D
  D\frac{\partial}{\partial D} + \gamma_\alpha \alpha
  \frac{\partial}{\partial \alpha } +\beta(u)\frac{\partial}{\partial
  u} \right)\notag\\& \times yf(D\kappa^2t, y\alpha \kappa^{-d/2},u)=0
\end{align}
where $\gamma_{\phi}=\kappa \partial_{\kappa} \ln Z_{\phi}$,
$\gamma_{D}=\kappa \partial_{\kappa} \ln Z_{D}$,
$\gamma_{\alpha}=\gamma_{\phi}/2$ and $\beta(u)=
\kappa\partial_{\kappa} u$, with $Z_{\phi}$, $Z_{D}$ the standard
renormalization factors \cite{Cardy99}.

Equation \eqref{eq:dim-ps} gives the following relations for the
differential operators
\begin{align}
 \kappa\frac{\partial}{\partial\kappa}&=2t\frac{\partial}{\partial
 t}-\frac{d}{2}\alpha\frac{\partial}{\partial\alpha}\ ,\nonumber\\
D\frac{\partial}{\partial D}&=t\frac{\partial}{\partial t}\ ,\label{eq:opid3}\\
\alpha\frac{\partial}{\partial\alpha}&=y \frac{\partial}{\partial y}
 -1 ,\nonumber
\end{align}
hence the generic solution at the fixed point is
\begin{align}\label{eq:gfscaling}
  \bar{F}(s,t)=t^{-\frac{\gamma^*_{\phi}}{2+\gamma^*_D}}y \bar{f}\left
  ( yt^{\frac{d/2-\gamma^*_{\phi}/2}{2+\gamma^*_D}} \right),
\end{align}
where starred parameters are values of corresponding $\gamma$ functions
at the fixed point.

Using the standard RG technique \cite{Amit84}, the function $\bar{f}$
 up to $O(\epsilon)$ and a multiplicative non-universal amplitude can
 be obtained  by replacing in Eq. \eqref{Eq:GF_one-loop} $u^2\to
 \epsilon/12$, $rt$ $\to$ $-C_y y
 t^{(d/2-\gamma^*_{\phi}/2)/(2+\gamma^*_D)}$ and $t/\bar t\to 1$;
 where $C_y>0$ is a second non-universal amplitude,
 $\gamma^*_{\phi} = -\epsilon/6$ and $\gamma^*_D = -\epsilon/12$ up to
 $O(\epsilon)$ \cite{Janssen05}.

Since the difference between $p(n,t)$ and $p(n+1,t)$ for large $t$  
is small, the generating function $\bar{F}$ can be approximated by a
Laplace transform with the parameter $\omega=-\ln s$. Applying the inverse
Laplace transform in which we approximate $y\approx -\omega$,  we obtain
\begin{equation}\label{eq:pnt-scaling}
  p(n,t)=t^{-d/z} \bar{g} \left(\frac{n}{ t^{d/(2z)+\eta/2}} \,
  \right)
\end{equation}
where we have introduced the exponents $z=2+\gamma^*_D$ and
$\eta=-\gamma^*_\phi/(2+\gamma^*_D)$ defined by scaling properties of the
two point Green function \cite{Cardy99}.  Equation   \eqref{eq:pnt-scaling}
along  with the relation $p_s(t)=\int dn p(n,t)$ gives the survival
exponent $\delta=(d/z-\eta)/2$,  which is the well-known hyperscaling
relation.

\begin{figure}
  \includegraphics[scale=0.7]{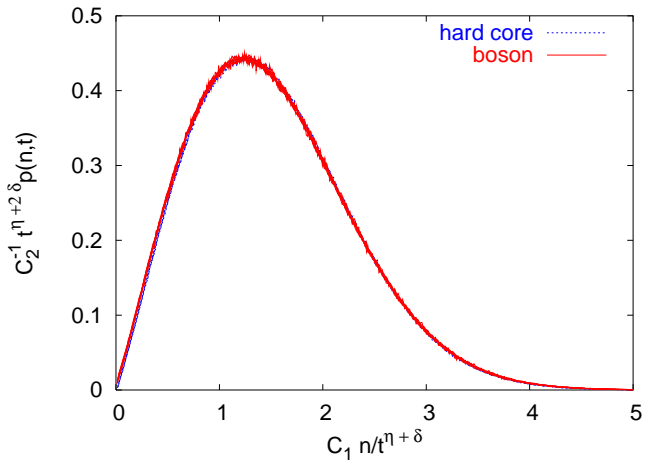} \caption{\label{fig:pnt} Critical
  scaling function for the PNPD in one dimension with full lines for
  boson models (pair annihilation and pair coagulation) and
  with dashed lines for hard-core particles models (the contact process
  and the bond directed percolation). The curve corresponding to the
  contact process is chosen for reference, and the other data sets
  are plotted for the best overlap with the reference curve by varying
  the constants $C_1$ and $C_2$.}
\end{figure}

The scaling analysis predicts that the critical PNPD at large $t$ has
the form
\begin{equation}
 p(n,t)=\frac{C_2}{t^{\eta+2\delta}}\;g\!\left (C_1\frac{
 n}{t^{\eta+\delta} }\right) \ ,
\label{Eq:cPNPD}
\end{equation}
where $C_1$, $C_2$ are two non-universal metric factors which depend on
the values of the microscopic parameters and $g$ is the universal
scaling function. 
The two-scale factor universality is worth while to be mentioned at this 
point.
Since we are interested in the dynamic behavior
of the system, there are, in general, three independent metric factors which
are related to the number of particles, the time, and the
distance from the critical point, respectively.
The third metric factor does not appear in Eq. (\ref{Eq:cPNPD}), because
the system is at the critical point. This should be compared to  
the two-scale factor universality in equilibrium \cite{Stauffer72}.

In one dimension, it is relatively easy to test
numerically the above scaling form.
We performed Monte Carlo simulations for the bosonic model
\cite{SCP04} described by Eq.~\eqref{eq:tran} at two points in the
parameter space $(D_l,\lambda_a, \lambda_b,\lambda_c)$ on the critical
manifold: $D_l=0.1 $ and $0.2$, $\lambda_a= (1-2D_l)p_c $,
$\lambda_b=(1-2D_l)(1-p_c) $, $\lambda_c=0.5 $ with $p_c= 0.11056(1)$
and $0.13345(4)$, respectively.  We measured the PNPD at two different
times $t=4000$ and $8000$ for $2\times 10^7$ samples.  As shown in
Fig.~\ref{fig:pnt}, we overlap the scaling functions almost perfectly
by varying the constants $C_1$ and $C_2$, which implies that the
scaling function $g$ is universal. The numerical values of the exponents $\eta$
and $\delta$ are taken from Ref.~\cite{Jensen99}:
$\eta=0.313686$ and $\delta=0.159464$. In the same figure, we
also plotted the scaling function of a variant model which has a pair
annihilation process $2A\to\varnothing$ instead of the pair
coagulation $2A\to A$.  This one also overlaps perfectly with the
others.

As pointed out in Introduction, there are many other models belonging
to the DP universality class. In the field theoretical formalism, they
differ by extra irrelevant coupling terms in the action besides the
terms in Eq.~\eqref{eq:continuumaction}.  Among these models, of a
particular interest are the hard-core models where the dynamic rules
accept only one or zero particle per site.  As such examples, we have
collected the PNPD data for the contact process \cite{Harris74} and
the directed bond percolation \cite{Hinrichsen00} in one dimension at
the same times as for the previous data.  Their scaling functions can
be also overlapped very well with those of the bosonic models in the
same manner. We have noticed a minute deviation between the two
scaling functions at small values of the scaling variable $C_1
n/t^{\eta+\delta}$, but a more detailed analysis shows that this
deviation is due to corrections to scaling which decay very slowly in
this region.

\section{\label{sec:concl}Conclusions}

We have shown that the generating function of the particle-number 
probability distribution of a DP model can be expressed
as a series of cumulants of the associated field theory. Using the
recurrence properties of the tree diagrams, we have calculated
explicitly the mean field solution and the logarithmic corrections
for the PNPD at the upper critical dimension.

From the renormalization group analysis, we have shown that the
scaling function for the PNPD is universal in the critical state and
the microscopic details of the model are included in two non-universal
metric factors.  We have tested the universality numerically in one
dimension for both boson and hard-core particle models. Numerical
simulations show that all models form the universality with only two
adjusting parameters in the critical state.
 
\begin{acknowledgments}
LA thanks  Hans-Karl Janssen and Frederic van Wijland  for illuminating
discussions and criticism on RG issues. This work was supported by the
European Community Marie Curie Fellowship scheme under contract
No. HPMF-CT-2002-01910.
\end{acknowledgments}

\appendix*

\section{\label{sec:a1} One-loop calculation and logarithmic corrections}

The logarithmic corrections to the critical scaling at the upper
critical dimension can be obtained using the RG
equation~\eqref{eq:callan}. For the calculation of logarithmic
correction of the critical GF up to the sub-leading order, we need the
expression of the critical GF up to one-loop order and the Wilson
functions up to two-loop order. The Wilson functions have been
calculated previously in Ref. \cite{Janssen00}, hence we have only to
calculate one-loop correction to the critical GF.

\begin{figure}[t]
\includegraphics[width=0.35\textwidth]{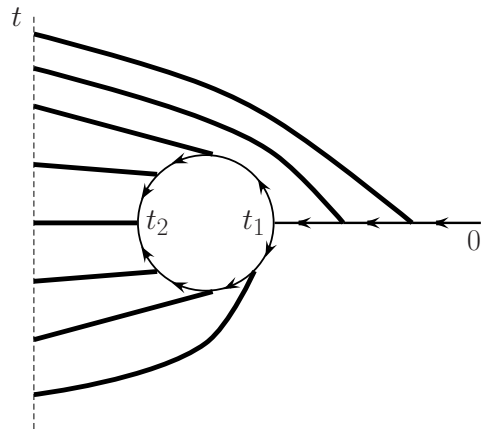}
\caption{\label{fig:loops} A generic  one-loop
  diagram. The thick line represents the MF solution of the GF whose
final time is $t$.}
\end{figure}

\begin{figure}[b]
\includegraphics[width=0.47\textwidth]{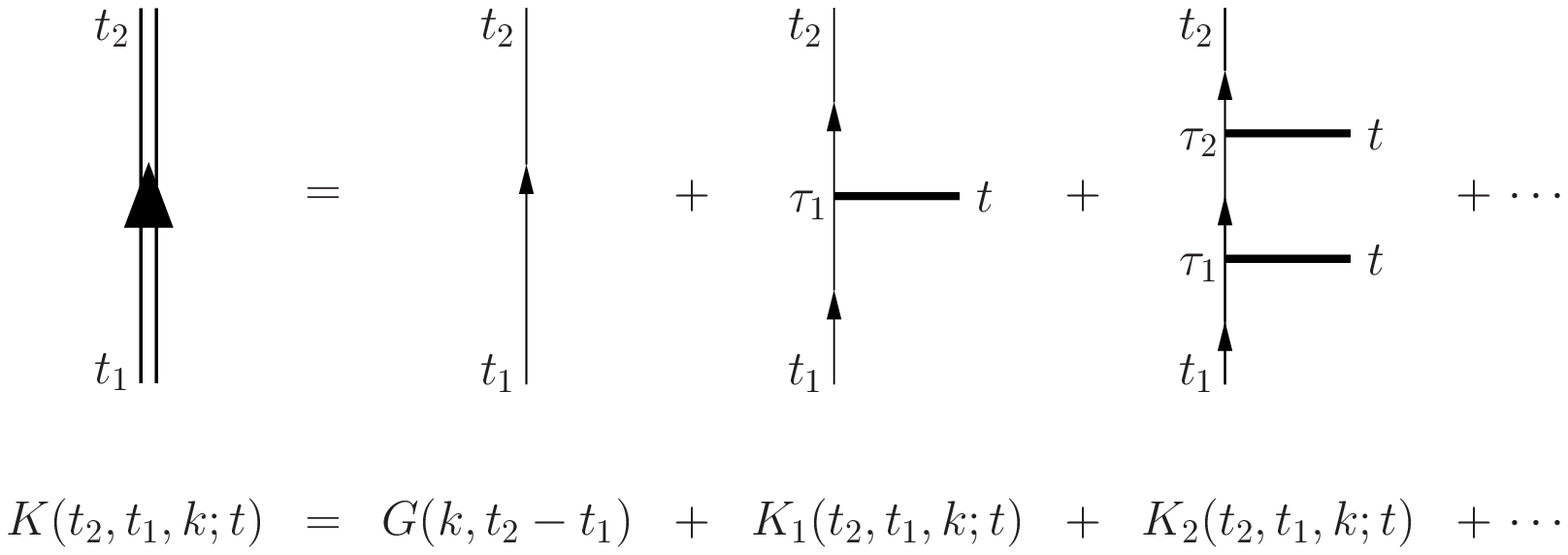}
\caption{\label{fig:dressed_tree} 
The one-particle Green function with insertions of trees.}
\end{figure}

The generic topological structure of a one-loop diagram is shown in
Fig.~\ref{fig:loops}. 
To sum up all such diagrams, we need the one-particle Green function
with insertions of trees whose end points are located at the same
final time $t$. The graphical 
representation of such calculations is shown in Fig. \ref{fig:dressed_tree}.

At zero mass, the one-particle Green function with 
insertions of $p$ trees is
\begin{align}
  & K_p(t_2,t_1,k;t) = \notag\\ &(\alpha g)^p\int_{t_1}^{t_2}
d\tau_1\int_{\tau_1}^{t_2} d\tau_2\dots \int_{\tau_{p-1}}^{t_2}d
\tau_p T(t-\tau_1)\dots T(t-\tau_p) \notag\\ &\times
e^{-Dk^2(t_2-t_1)} \notag \\ &=
\frac{2^p}{p!}\left[\log\left(\frac{1+r(t-t_2)}{1+r(t-t_1)}
\right)\right]^p e^{-Dk^2(t_2-t_1)} \ ,
\end{align}
where $t_2>t_1$ is assumed and
\begin{align} \label{eq:T}
T(t)=(s-1)/(1+r t) 
\end{align} 
is the MF solution of the critical GF, i.e., $\bar F(s,t)$ at $\mu=0$
and $r=-(s-1)\alpha g/2$. Hence the sum over all tree insertions is
\begin{align}
&K(t_2,t_1,k;t)=\sum_{p=0}K_p(t_2,t_1,k;t) \notag \\ &=
\theta(t_2-t_1) \left(\frac{1+r(t-t_2)}{1+r(t-t_1)}\right)^2
e^{-Dk^2(t_2-t_1)} \ .
\end{align}
 
\begin{figure}
\includegraphics[width=0.47\textwidth]{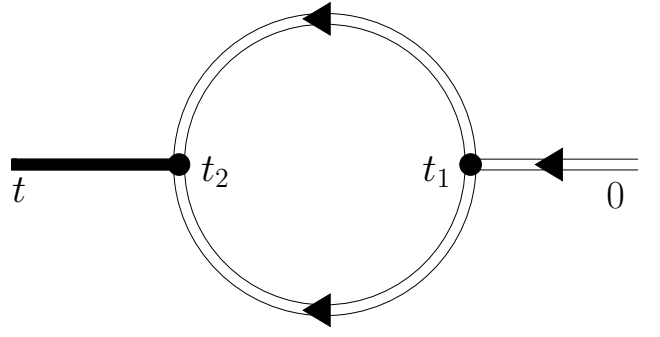}
\caption{\label{fig:GF_oneloop} 
The diagram of one-loop corrections to the GF. The thick line whose
terminal point is $t$ is the MF GF and the double-line with arrow
is the one-particle Green function with insertions of all trees.}
\end{figure}

With the above Green function the sum over all 
one-loop diagrams has the form (see also Fig. \ref{fig:GF_oneloop}).
\begin{align}
  L=&-\frac{1}{2}\frac{g^2}{ (8D\pi)^{d/2}} \int_{0}^{t}
   dt_1\int_{t_1}^t dt_2 T(t-t_2) \notag \\ &\times
   K^2(t_2,t_1,0;t)K(t_1,0,0;t)(t_2-t_1)^{-d/2} \ ,
\end{align}
where the integral over momentum was done.
The integrals in the foregoing expression are standard and we have,
using dimensional regularization,
\begin{align}
  L&=-\frac{u^2 (s-1)}{(1+r t)^2}\left[ -\frac{1}{1-\epsilon/2}\left(
  \frac{2}{\epsilon} \left(\frac{t}{\bar
  t}\right)^{\frac{\epsilon}{2}} + rt\left( \frac{t}{\bar t}
  \right)^{\frac{\epsilon}{2}} \right) \right . \notag \\
  &-3\frac{2}{\epsilon}\frac{rt}{1+\epsilon/2} \left(\frac{t}{\bar t}
  \right)^{\frac{\epsilon}{2}} 
  +3\frac{1}{1+\epsilon/2}\left(rt-\log(1+rt)\right) \notag \\
  &\left. -\frac{1}{2+\epsilon/2}\left(
  2rt-2\log(1+rt)-\frac{r^2t^2}{1+rt} \right) +O(\epsilon)\right],
\label{eq:oneloop-eps}
\end{align}
where $\epsilon=4-d$, $u=\sqrt{K_4}(g/D)\kappa ^{-\epsilon/2}/4$ with
$K_4 = 1/(8 \pi^2)$ is the dimensionless coupling constant, and ${\bar
t}^{-1}=8\pi D\kappa^2$ with $\kappa^{-1}$ an arbitrary length scale.
The renormalized GF can be obtained using the standard
renormalization constants $Z_{\phi}=1+(2/\epsilon)u^2$ and
$Z_g=1+(5/\epsilon)u^2$; see Eq.  \eqref{eq:cumexp-aplitudes}.
Hence, the renormalized GF up to one-loop order is
\begin{align}
  \bar F(s,t)&=\frac{(s-1)}{1+r t}\left[1+\frac{u^2}{1+r t} \biggl(
   1+\log(t/\bar t) \right.  \notag \\ & \left.  +2\log(1+rt)+rt\biggl(
   3\log(t/\bar t) -\frac{4+\frac{9}{2}rt}{1+rt} \biggr) \biggr) \right] \ .
\label{Eq:GF_one-loop}
\end{align}

For the calculation of the logarithmic correction we use the Wilson
functions in two-loop order from Ref. \cite{Janssen00}
 \begin{align}
   \beta(u)&=6u^3-\frac{1}{4}\left(
   169+106\ln\frac{4}{3} \right)u^5, \label{eq:beta}\\
   \gamma_\phi(u)&=-2 u^2+2\left(6-9\ln\frac{4}{3}
   \right)u^4,\\
   \gamma_D(u)&=-u^2+\frac{1}{4} \left(
   17-2\ln\frac{4}{3} \right)u^4, \label{eq:will2}
 \end{align}
with the observation that in Ref.~\cite{Janssen00} the Wilson functions
are written for the square of our coupling constant multiplied by
a different factor.

The characteristic system to be solved is (see
Eqs. (\ref{eq:callan}-\ref{eq:opid3}))
\begin{align}
  \rho\frac{du}{d\rho}&=\beta(u),\\
 \rho\frac{d\alpha(\rho)}{d\rho}&=(-2+\gamma_\phi(u)/2)\alpha(\rho),\\
 \rho\frac{d t(\rho)}{d\rho}&=(2+\gamma_D(u))t(\rho) \ .
\end{align}

Using $d\rho/\rho=dw/\beta(w)$, we have
\begin{align}
  &t(w)=t C_t w^{a_0}\exp[-1/(\beta_3 w^2)+b_0w^2\!+O(w^4)]\label{eq:tw}
  \ ,\\ &\alpha(w)=C_\alpha w^{a_1}\exp[1/(\beta_3 w^2) +b_1 w^2
  +O(w^4)] \ ,\\ & \int_u^{w} \frac{\gamma_\phi(w')}{\beta(w')}dw' =
  N'w^{a_2}\exp[b_2 w^2+O(w^4)]\ ,
\end{align}
where $C_t$, $C_\alpha$ are determined from the initial condition
$t(w=u)=t$, $\alpha(w=u)=\alpha$. 
We denote by $\beta_i$, $\phi_i$ and
$D_i$ the coefficients in the expansion of the Wilson functions
$\beta(w)$, $\gamma_\phi(w)$ and $\gamma_D(w)$ that generically can be
written $Q=\sum_i c_{Q,i} u^i$, similar to the notations used in
Ref. \cite{Janssen04}. 
The constants $a_0$, $a_1$, $a_2$, $b_0$,
$b_1$, $b_2$ are defined by the following relations
\begin{align}
 a_0&=\frac{1}{\beta_{3}^2}(D_2\beta_3-2\beta_5)\ , \label{eq:a0}\\
  a_1&= \frac{1}{2\beta_{3}^2}(\phi_2\beta_3+4\beta_5)\ ,\\
  a_2&=\phi_2/\beta_{3}\ , \\ 
 b_0&=\frac{1}{2\beta_{3}^{2}}\left(D_4\beta_3 -D_2
  \beta_5\right)+\frac{\beta_{5}^2}{\beta_{3}^3}\ , \\
b_1&=
  \frac{1}{4\beta_{3}^2} \left( \phi_4\beta_3 -
  \phi_2\beta_5\right)-\frac{\beta_{5}^2}{\beta_{3}^3}\ , \\
  b_2&=\frac{1}{2\beta_{3}^2}(\phi_4\beta_3-\phi_2\beta_5)\
  . \label{eq:b2}
\end{align}
Under scaling the generating function transforms as follows:
\begin{align}
 \bar{F}(s,t;w, \alpha)& =\exp\left[\int_{u}^wdw'
  \gamma_{\phi}(w')/\beta(w')\right] \notag\\ &\times\bar{F}(s,t(w);
  w, \alpha(w))\ ,
\end{align}
where we can eliminate $w$ with the condition $t(w)=X_0
\bar t$ with $X_0$ an arbitrary constant of order one. The GF at large
$t$ is 
\begin{widetext}
\begin{align}\label{eq:gflog}
 \bar F(s,t)&=\frac{N'(s-1) w^{a_2}e^{b_2w^2+O(w^4)}(1+(\ln X_0+1)w^2)}
  {S(w,s,t)} \left[1+\frac{w^2}{S(w,s,t)} \biggr( 1+\log X_0
  +2\log(S(w,s,t))\right.  \notag \\ & \left. -(1-S(w,s,t))\left(3\log
  X_0 -\frac{9}{2} +\frac{1}{2S(w,s,t)}\right) \biggr) \right]\
  ,\end{align} where
\begin{align}
S(w,s,t)=1-\frac{1}{2} (s-1) \bar{D} t w^{a_0+a_1+1}
  e^{(b_0+b_1) w^2+O(w^4)} \ ,
\end{align}
$\bar{D}=C_\alpha C_t D K_4^{-1/2}$ and for large $t$ we have from
Eq. \eqref{eq:tw}
\begin{align}
  w&=\tau^{-1/2}\exp\left[ \frac{a_0\beta_3}{4}\frac{\ln \tau}{\tau} 
+O\left (\tau^{-2}\ln^2\tau \right ) \right] \ ,\label{eq:woftau} \\ 
  \tau&=\beta_3\ln(t/t_0)\label{eq:tauoft} \ .
\end{align} with  $t_0= (X_0/C_t) \bar{t}$.
\end{widetext}

\end{document}